\documentclass[pra,twocolumn,showpacs,preprintnumbers,amsmath,amssymb,floatfix,superscriptaddress]{revtex4-1}

\usepackage{graphicx}
\usepackage{dcolumn}
\usepackage{bm,bbm}

\begin{document}


\title{Adiabatic passage in photon-echo quantum memories}

\author{Gabor Demeter}
\email{demeter.gabor@wigner.mta.hu}
\affiliation{Wigner Research Center for Physics, Hungarian Academy of Sciences, Konkoly-Thege
Mikl\'os \'ut 29-33, H-1121 Budapest, Hungary}

\date{\today}

\begin{abstract}
Photon-echo based quantum memories use inhomogeneously broadened, optically thick ensembles of
absorbers to store a weak optical signal and employ various protocols to rephase the atomic
coherences for information retrieval. 
We study the application of two consecutive, frequency-chirped control
pulses for coherence rephasing in an ensemble with a 'natural' inhomogeneous broadening.
Although propagation effects distort the two control pulses differently, chirped pulses that
drive adiabatic passage can rephase atomic coherences in an optically thick storage medium. 
Combined with spatial phase mismatching techniques to prevent primary echo emission, 
coherences can be rephased around the ground state to achieve secondary echo emission with close to
unit efficiency. Potential advantages over similar schemes working with $\pi$-pulses include
greater potential signal fidelity, reduced noise due to spontaneous emission and better capability
for the storage of multiple memory channels. 

\end{abstract}

\maketitle

\section{Introduction}
\label{intro}

Building a quantum memory for light is vital for creating future large scale
quantum communication networks and essential for several devices in quantum
information processing. We must store the quantum state of light in some material device
and be able to retrieve it efficiently and faithfully.
Thus intense research is going on to create an optical memory 
that could work right down to the single photon level,
using several different approaches \cite{Lvovsky2009,Simon2010}. 
In particular, photon-echo based techniques have been investigated extensively 
\cite{Tittel2010}.  

The first essential ingredient for any optical memory based on the photon-echo principle is an
inhomogeneously broadened ensemble of 'atoms' that absorb the signal and dephase for
storage.
Rare-earth ion dopants embedded in solid state lattices are a popular choice because they have
very long coherence times at low temperatures, the density of absorbers can be very large and
decoherence due to atomic motion is absent.
The second essential ingredient is a protocol to collectively rephase the atomic coherences of the
ensemble for the retrieval of the signal echo. 
The numerous techniques can be categorized in two wide groups. The
first one uses an atomic ensemble with a 'natural' inhomogeneous broadening and one or more strong
control pulses to rephase the coherences in the spirit of the
classical photon-echo phenomenon \cite{AllenEberly}. The second group uses specially
prepared atomic ensembles, whose absorption line shapes are crafted prior to signal
absorption.

The simplest technique of the first category is the
classical two-pulse photon-echo (2PE). It uses a single short $\pi$-pulse 
to rephase the coherences and does not require any initial state preparation
of the ensemble.
However, Ruggiero and coworkers showed \cite{Ruggiero2009} that it is unsuitable for a
quantum memory protocol for several reasons. First, rephasing
occurs when the ensemble is inverted, severely limiting the signal to noise ratio during
quantum state retrieval. Second, the control pulse is distorted during propagation, its bandwidth
decreases gradually and it develops a long tail that may interfere with the detection of the echo
\cite{Ruggiero2010}.
Furthermore, the protocol is extremely sensitive to the precise preparation of the control pulse,
as the pulse area of $\pi$ is in fact an unstable solution of the famed area equation.
Third, a control pulse whose bandwidth is wide enough to rephase the coherences
of the ensemble must be very short, with a high peak intensity, which may well
exceed the damage threshold in a crystal.

Noise from an inverted storage medium prevents quantum information storage in other 
cases as well \cite{Sangouard2010}, so techniques were proposed
to prevent the emission of the first echo and use a second control pulse to rephase the coherences
again. This secondary echo (in fact an echo of the primary echo) is emitted when
the atomic dipoles rephase around the ground state. Damon and coworkers \cite{Damon2011} used the
fact that if signal and control
pulses propagate in different directions, the primary echo fails the phase matching condition, so
it is silenced - a technique also employed in \cite{Ham2012}.
Another protocol \cite{Moiseev2011} uses a third atomic level and
strong Raman type interaction to store the signal in the coherences between
the two stable states. It employs special writing, rephasing and reading pulses to achieve rephasing
around the ground state.
An auxiliary electrical field gradient that broadens the absorption
line during the first rephasing can also be used to silence the primary echo \cite{McAuslan2011}. 

Techniques belonging to the second group achieve coherence rephasing around the
ground state
by preparing a special absorption feature in the storage medium. Controlled reversible
inhomogeneous broadening (CRIB) \cite{Kraus2006,Sangouard2007,Longdell2008,Tittel2010}  and gradient
echo memory (GEM) \cite{Moiseev2008,Hetet2008} techniques use a narrow absorption line
broadened by an externally applied inhomogeneous field. Reversing the field gradient
rephases the atoms, so inverting them is not necessary. These techniques have been
demonstrated to
work in solid state media \cite{Alexander2006,Staudt2007,Hedges2010} and used in more elaborate
configurations such as information storage in Raman coherences \cite{Carreno2011,Hosseini2012} or
polarization state qubit storage in three-level systems \cite{Viscor2011,Viscor2011a}.
Another technique is to craft an absorption feature composed of narrow, equidistant
peaks termed atomic frequency combs (AFC) \cite{Afzelius2009,DeRiedmatten2008,Minar2010}. Atomic
coherences then spontaneously rephase periodically, with a period given by the frequency spacing of
the peaks. The greatest difficulty with these techniques is the preparation of the required
absorption feature with sufficient optical depth.

As for techniques of the first category that use an unmanipulated absorption line, silencing the
primary echo still does not solve problems associated with control pulse propagation 
 in an optically dense medium, such as pulse distortion, high peak intensity and sensitivity to
the precise pulse area.
Recently, frequency-chirped control pulses that drive adiabatic passage (AP) between the atomic
states were proposed for use in photon-echo quantum memories
\cite{Minar2010,Damon2011,Pascual-Winter2013}. It has been shown, that while AP with a single
chirped pulse cannot, in general, rephase the coherences collectively, a pair of consecutive APs 
can under certain conditions, most notably when the control pulses are identical. With chirped
control pulses, the precise pulse area is not important and they can invert the same frequency range
of the atomic ensemble using much smaller peak intensities than $\pi$-pulses. 
For this reason, AP demonstrates superior performance compared to $\pi$ pulses also in rephasing
coherences in EIT based quantum memory experiments \cite{Mieth2012}.
However, the question of
pulse propagation effects remains. Even if the control pulses are identical at the entry of the
medium, they will surely be different at finite optical depths, because the second one propagates
in a gain medium inverted by the first one. For a collective
rephasing of the coherences, it is not only population transfer that
counts, but also the time integral of the adiabatic eigenvalues 
\cite{Minar2010,Pascual-Winter2013}. So how do pulse propagation effects modify the ability 
of a pair chirped control pulses to rephase atomic coherences?

In this paper we investigate the propagation properties of two consecutive, frequency-chirped
control pulses in an optically thick, inhomogeneously broadened atomic
ensemble. Calculating the distortion that the control pulses undergo, we investigate their ability
to collectively rephase the coherences of the ensemble. We also compare their
performance to that
of a pair of $\pi$ control pulses. We show that chirped control pulses are much more suitable for
rephasing an optically thick storage medium for multiple reasons. Finally, we calculate
the echo of a series of weak signal pulses and characterize the efficiency and fidelity of an
optical memory with chirped control pulses. We prove that together with phase mismatching to
extinguish the primary echo, frequency chirped
control pulses can be used effectively in optical quantum memories.

\section{Basic principles}

\begin{figure}[htb]
\includegraphics[width=0.45\textwidth]{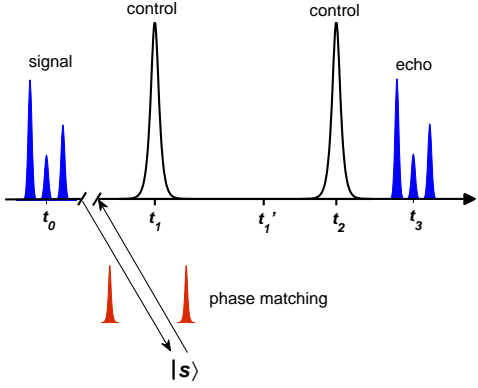}
\caption{Interaction timeline. A signal field at $t_0$ is followed by two consecutive control
pulses at $t_1$ and $t_2$, with the echo emission occurring at $t_3$. The primary echo at $t_1'$ is
silenced by spatial phase mismatching. In a variant of the protocol, a pair of counterpropagating
pulses can be used to transfer
atomic populations between the excited state and a third, stable state to achieve longer storage
times and obtain phase matching for backward echo emission.}
\label{fig_timeline}       
\end{figure}

We consider two variants of a photon-echo memory protocol in which an unmanipulated, 'natural'
inhomogeneously broadened absorption line is used for storage, and two consecutive control pulses
drive AP in
the ensemble twice to rephase the coherences around the ground state. The general timelines of the
variants are depicted in Fig. \ref{fig_timeline}. In the first one, we simply use an ensemble of
two-level atoms and two control pulses. In the second variant, we assume that an additional pair of
counterpropagating pulses transfer the excited state population to a third, long-lived state
$|s\rangle$ just after signal absorption 
as in several other protocols (e.g. \cite{Minar2010}). This step can extend storage time and perform
phase matching to enable backward echo emission. We envision a solid state medium where
inhomogeneous broadening is independent of the pulse propagation direction, and 
assume that $L\gg\lambda$ is fulfilled
for the length $L$ of the storage medium, so the primary echo can be silenced using
spatial phase mismatching \cite{Damon2011}.
We restrict our consideration
to signal and control pulse propagation along a single dimension. The reason is that the
interaction region where the signal is absorbed in photon-echo memory experiments is usually highly
elongated, so achieving AP with control pulses at an angle would probably require pulses with
prohibitively large intensities and/or very oblique beam shapes.  Finally, we assume that the signal
field is so weak (a few photon pulse) that it does not, in any way, interfere with control pulse
propagation, i.e. this can be computed in the 'empty' medium and the results
then used to calculate the triggering of echoes.

\subsection{Basic equations}

We consider propagation along a single direction and write the electric
field as a sum of forward and backward propagating modes, so the (classical) electric field
is:
\begin{equation*}
E(z,t)=\frac{1}{2}\left(E_f(z,t)e^{ikz-\omega_0t}+E_b(z,t)e^{-ikz-\omega_0t}+c.c.\right)         
\end{equation*}
with the slowly varying envelope functions $E_f(z,t)$ and $E_b(z,t)$. Here $\omega_0$ is the
central frequency of the inhomogeneously broadened absorption line and we use the time-dependent
complex phase of the envelope functions to include detunings and frequency-modulations in our
description. 

We use the rotating frame Hamiltonian $\hat{H}_a=\hbar\Delta|e\rangle\langle e|$ to 
describe a two-level system with transition frequency
$\omega_{eg}=\omega_0+\Delta$, offset
by $\Delta$ from the inhomogeneously broadened line center. In addition, we use the standard 
dipole interaction Hamiltonian and the rotating wave approximation. Thus we obtain the following
equations for the probability
amplitudes $\alpha(t;z,\Delta),\beta(t;z,\Delta)$ that describe the state of an atom at point $z$
as $|\psi\rangle=\alpha(t;z,\Delta)|g\rangle+\beta(t;z,\Delta)|e\rangle$:
\begin{eqnarray*}
\partial_t\alpha(t;z,\Delta)&=&\frac{i}{2}\left(\Omega_f^*(t,z)e^{-ikz}
+\Omega_b^*(t,z)e^{ikz}\right)\beta(t;z,\Delta)\nonumber\\
\partial_t\beta(t;z,\Delta)&=&\frac{i}{2}\left(\Omega_f(t,z)e^{ikz}
+\Omega_b(t,z)e^ {-ikz}\right)\alpha(t;z,\Delta)\nonumber\\
&&-i\Delta\beta(t;z,\Delta)
\end{eqnarray*}
Here $\Omega_{f,b}=d E_{f,b}/\hbar$ are the Rabi frequencies of the
forward and backward propagating fields with $d=\langle e|\hat{d}|g\rangle$ the dipole matrix
element. We have neglected all decay processes in this description, so the overall interaction
time must be much shorter than any atomic population or coherence decay times.
It is convenient to decompose the probability amplitudes as a series
of spatial Fourier modes: 
\begin{eqnarray}
\alpha(t;z,\Delta)&=&\sum_n\alpha_n(t;z,\Delta)e^{inkz}\nonumber\\
\beta(t;z,\Delta)&=&\sum_n\beta_n(t;z,\Delta)e^{inkz}\nonumber
\end{eqnarray}
$\alpha_n(t;z,\Delta)$ and $\beta_n(t;z,\Delta)$ 
still depend on $z$, but now vary only slowly on the scale of the light wavelength, similar to
$\Omega_{f,b}$. Using $kL\gg1$, we can
separate the evolution equation for the slowly varying probability amplitudes:
\begin{eqnarray}
\partial_t\alpha_n&=&\frac{i}{2}\left(\Omega_f^*\beta_{n+1}
+\Omega_b^*\beta_{n-1}\right)\nonumber\\
\partial_t\beta_n&=&\frac{i}{2}\left(\Omega_f\alpha_{n-1}
+\Omega_b\alpha_{n+1}\right)-i\Delta\beta_n
\label{eq_bloch1}
\end{eqnarray}
(The explicit dependence on $t$, $z$ and $\Delta$ has been suppressed for brevity.)

To obtain the spatiotemporal evolution of the fields from the wave equation, we employ the
slowly varying envelope approximation. Using $kL\gg1$, the equations for $\Omega_{f}(z,t)$ and
$\Omega_{b}(z,t)$ can be separated:
\begin{eqnarray}
\left(\frac{1}{c}\partial_t+\partial_z\right)\Omega_f(t,z)&=&
i\frac{\alpha_d}{\pi g(0)}\mathcal{P}_1(z,t)\nonumber\\
\left(\frac{1}{c}\partial_t-\partial_z\right)\Omega_b(t,z)&=&
i\frac{\alpha_d}{\pi g(0)}\mathcal{P}_{-1}(z,t)
\label{eq_maxwell1}
\end{eqnarray}
Here $g(\Delta)$ is the inhomogeneous line shape function,  
$\alpha_d=\pi g(0)k\mathcal{N}d^2/\varepsilon_0\hbar$ is the absorption constant and 
we have introduced the notation
\begin{equation}
 \mathcal{P}_{\pm1}(z,t)=\int\sum_n\alpha_n^*\beta_{n\pm1}g(\Delta)d\Delta
\label{eq_polarization}
\end{equation}
for the forward and backward parts of the polarization. The fact that each field interacts
only with the corresponding part of the polarization is an expression of the spatial phase matching
condition.
Eqs. \ref{eq_bloch1} and \ref{eq_maxwell1}, together with \ref{eq_polarization} constitute the set
of Maxwell-Bloch equations for our case. They can be solved analytically for the signal field
in the weak excitation limit \cite{Ruggiero2009}, but can only be solved numerically for the control
pulses
and for the echo. However, we assume that the pulses propagate in complete time separation, so the
solution is somewhat simplified - during the time interval $[t_i-T,t_i+T]$ where the $i$-th pulse
has a finite amplitude, it is enough to solve Eqs. \ref{eq_bloch1} for one or two pairs of
 amplitudes $\{\alpha_n,\beta_{n\pm1}\}$ that $\Omega_i$ couples, those that may be nonzero at
the time of the $i$-th pulse and may contribute to $ \mathcal{P}_{\pm1}$. In the second variant of
the protocol
where two additional pulses transfer the atomic excitation between $|e\rangle$ and a third state
$|s\rangle$, we simply assume that they are perfect $\pi$-pulses or a pair of identical chirped
pulses. Because the $|e\rangle\leftrightarrow|s\rangle$ transition is virtually empty in the case
of weak signal fields, the medium is perfectly transparent for these pulses, propagation effects
need not be taken into account for them.

\subsection{Primary echo suppression via spatial phase mismatching}
\label{sec_protocols}
\begin{figure}
\includegraphics[width=0.45\textwidth]{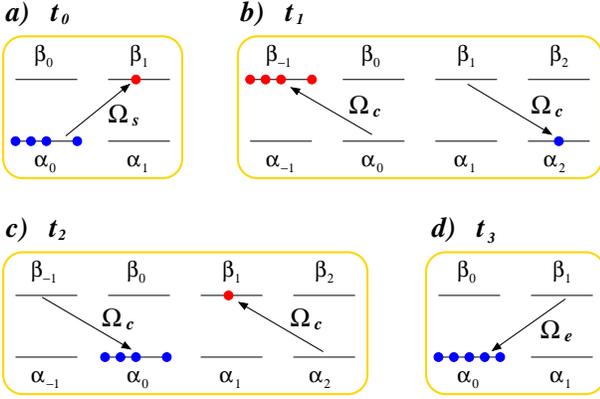}
\caption{Steps of the first memory variant showing the populated probability amplitudes and the
transfers driven by the various fields. Two backward propagating control pulses rephase the
coherences around the ground state and a forward echo is emitted.}
\label{fig_scheme1}       
\end{figure}

\begin{figure}
\includegraphics[width=0.45\textwidth]{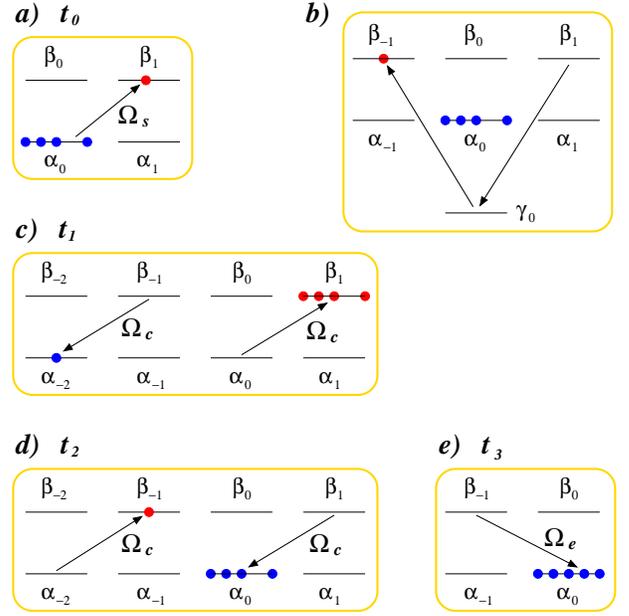}
\caption{Steps of the second memory variant showing the populated probability amplitudes and the
transfers driven by the various fields. A pair of counterpropagating pulses on the
$|e\rangle\leftrightarrow|s\rangle$ transition and two forward propagating control pulses
rephase the coherences around the ground state and a backward echo is emitted.}
\label{fig_scheme2}       
\end{figure}

The main steps of the two variants are sketched 
in Figs. \ref{fig_scheme1} and \ref{fig_scheme2}, which depict the various probability amplitudes
that differ from zero at certain times.
Assuming that we start with a spatially
homogeneous medium, initially only $\alpha_0$ is nonzero.
In the first case, the absorption of the forward propagating signal pulse at $t_0$ 
creates coherences in the $\{\alpha_0,\beta_1\}$ amplitude pair 
[Fig.\ref{fig_scheme1} a)].  The first control pulse at $t_1$, which propagates in the backward
direction, inverts the atoms, transferring the populations to $\beta_{-1}$ and $\alpha_2$
[Fig. \ref{fig_scheme1} b)]. 
When coherences rephase around the excited state at $t_1'$  the polarizations $\mathcal{P}_1$
and $\mathcal{P}_{-1}$ are both zero - indeed only $\mathcal{P}_{-3}$ is nonzero - so the primary
echo is silenced.
At $t_2$ the second control pulse
(backward propagating) returns the populations to $\{\alpha_0,\beta_1\}$ [Fig.\ref{fig_scheme1} c)],
so the rephasing at $t_3$ occurs around the ground state, 
giving rise to $\mathcal{P}_1$, i.e. forward echo emission [Fig.\ref{fig_scheme1} d)]. 

The first step of the second variant is identical to the first one [Fig. \ref{fig_scheme2} a)],
but now it is followed by a transfer of the excited state population to the shelving state
$|s\rangle$ by a forward propagating pulse [Fig. \ref{fig_scheme2} b)]. The dephasing is
halted and the signal
stored in the coherence between $\{\alpha_0,\gamma_0\}$ as in other protocols
\cite{Minar2010,Moiseev2011}. Upon demand a second pulse, this time backward propagating,
transfers the population from $\gamma_0$ to $\beta_{-1}$ 
and rephasing can proceed. We assume that the difference between the wavelengths of the control
pulse pair driving the shelving transition $|e\rangle\leftrightarrow|s\rangle$ and the signal field
is small ($|k_c-k_s|L\ll1$), so this pulse pair also performs the necessary phase matching required
for backward
echo emission. Because this transition is virtually empty, we
simply assume them to be a pair of identical chirped pulses and need not
consider their propagation. 
Next, the first control pulse at $t_1$, this time forward propagating, inverts the atoms,
transferring the populations to $\beta_{1}$ and $\alpha_{-2}$ [Fig. \ref{fig_scheme2} c)].
When the coherences rephase around the excited state at $t_1'$, only $\mathcal{P}_{3}$ is nonzero.
Rephasing occurs at $t_3$, after the second control pulse [Fig. \ref{fig_scheme2} d)] 
giving rise to $\mathcal{P}_{-1}$, i.e. a backward echo is generated [Fig.\ref{fig_scheme2} e)].
Note however, that because of possible imperfections in the population transfer process, we must in
fact use more amplitude pairs than depicted in Figs. \ref{fig_scheme1} and \ref{fig_scheme2} when
computing echo emission numerically.

\section{Coherence rephasing with adiabatic passage}
\label{sec_rephasing}
\subsection{Properties of the time evolution operator}

The control pulses at $t_1$ and $t_2$ must be able to rephase a sufficiently large region
of the atomic ensemble in terms of optical depth and frequency range in order to 
trigger echo emission with high efficiency and good fidelity.
To investigate whether the coherences imprinted by the signal  
can be rephased by the pulses, we construct the time evolution operator $\hat{U}(\Delta,z)$ that
connects
the values of a pair of probability amplitudes at $t=t_3-T$ just before
echo emission with their values at $t=t_0+T$
just after the signal pulse has been absorbed:
\begin{equation*}
\left(\begin{array}{c}\alpha_n'\\ \beta_{n\pm1}'\end{array}\right)=
\hat{U}(\Delta,z)\left(\begin{array}{c}\alpha_n\\ \beta_{n\pm1}\end{array}\right).
\end{equation*}
(The upper sign in $\beta_{n\pm1}$ is valid for forward propagating control pulses, while
the lower sign for backward ones.)
$\hat{U}(\Delta,z)$ can be
constructed from the time evolution matrices $\hat{U}^{C1}(\Delta,z)$ and
$\hat{U}^{C2}(\Delta,z)$ of the two control pulses that propagate the amplitudes during the
time intervals $[t_{1,2}-T',t_{1,2}+T']$ and the free evolution matrices between the various pulses.
(See the appendix for a short derivation, or \cite{Pascual-Winter2013} for a detailed treatment.)

Let us now define the quantities $\mathcal{R}^{C1}$ and $\mathcal{R}^{C2}$ using the off diagonal
matrix elements of $\hat{U}^{Cj}(\Delta,z), j\in\{1,2\}$:
\begin{equation*}
 \mathcal{R}^{Cj}(\Delta,z)=\left[\hat{U}^{Cj}(\Delta,z)\right]_{12}
\cdot\left[\hat{U}^{Cj}(\Delta,z)\right]_{21}^*
\end{equation*}
Clearly, $\mathcal{R}^{Cj}(\Delta,z)$ is the quantity that is relevant for the collective rephasing
of coherences by the $j$-th control pulse. First, its magnitude gives the probability that the
control pulse inverts the atomic states. Second, $|\mathcal{R}^{Cj}(\Delta,z)|=1$ implies 
$\left[\hat{U}^{Cj}(\Delta,z)\right]_{11}=\left[\hat{U}^{Cj}(\Delta,z)\right]_{22}=0$, so
in this case the atomic
coherences are transformed by the pulse during the time interval $[t_j-T',t_j+T']$ as
\begin{equation*}
(\alpha_n^*\beta_{n\pm1})'=\left(\mathcal{R}^{Cj}(\Delta,z)\alpha_n^*\beta_{n\pm1}\right)^*. 
\end{equation*}
For a perfect $\pi$-pulse, $\mathcal{R}^{Cj}=1$, while for a control pulse that creates AP between
the two atomic states
\begin{equation}
 \mathcal{R}^{Cj}(\Delta,z)=
-e^{i[\Lambda_j^-+\Phi_j(t_j-T')]}\cdot e^{i[-\Lambda_j^++\Phi_j(t_j+T')]}
\end{equation}
(see Eq. \ref{UCP1}). Here $\Lambda_j^\pm$ are the time integrals of the adiabatic
eigenvalues for the duration of the control pulse
which depend explicitly on $\Delta$ and, through the complex pulse amplitude
$\Omega_{j}(z,t)$ which changes as the control pulse propagates, also on $z$. $\Phi_j(t)$ is the
complex phase of $\Omega_{j}(z,t)$. 
In general, a single control pulse is able to collectively rephase the coherences
in some region
of the ensemble if, in this domain of $\Delta$ and $z$ both $|\mathcal{R}^{Cj}(\Delta,z)|=1$
and $arg[\mathcal{R}^{Cj}(\Delta,z)]=\mathrm{const.}$ are satisfied simultaneously.
This is usually not the case. 
(Rephasing is
 possible when the control pulse amplitude is so large such that the dependence of
$\Lambda_j^\pm$ on $\Delta$ is negligible, but this presents the same problems with peak
intensity as a short $\pi$-pulse.)

When two chirped control pulses are used in succession for rephasing, both of which create AP, the
overall transformation of the atomic coherences becomes:
\begin{multline}
 (\alpha_n^*\beta_{n\pm1})'=(\alpha_n^*\beta_{n+1})
\cdot\left[\hat{U}(\Delta,z)\right]_{11}^*
\cdot\left[\hat{U}(\Delta,z)\right]_{22}\\
=(\alpha_n^*\beta_{n\pm1}) 
\mathcal{R}^{C1}(\Delta,z)[\mathcal{R}^{C2}(\Delta,z)]^*
e^{i\Delta(2t_2-2t_1+t_0-t_3+2T)}\\
\label{eq_rephase}
\end{multline}
(see Eqs. \ref{Umatrix} - we stress again, that this formula is valid only when both pulses create
AP). From Eq. \ref{eq_rephase} it is clear that a pair of chirped control pulses can rephase
atomic coherences collectively even if a single one cannot \cite{Minar2010,Pascual-Winter2013}. If
$|\mathcal{R}^{C1}|=|\mathcal{R}^{C2}|=1$ and $arg(\mathcal{R}^{C1})=arg(\mathcal{R}^{C2})+2m\pi$
are
both satisfied simultaneously, coherences will
be just prepared for rephasing at $t_3$ by the control pulses provided that $2t_2-2t_1+t_0-t_3=0$.
At the entry of the storage medium, this can easily be achieved by the use of two identical control
pulses. But the two control pulses will be deformed during propagation in a different way, because
they experience different initial conditions. The first pulse is absorbed, while the second one
propagates through an inverted medium and is thus amplified. Thus we must also investigate just
how fast propagation effects destroy the capability of the control pulse pair to rephase.

\subsection{Simulation results}

To investigate whether a pair of control pulses would be able to rephase an optically
thick ensemble of two-level atoms, we solved Eqs. \ref{eq_bloch1}, \ref{eq_maxwell1} and
\ref{eq_polarization} for a pair of propagating chirped control pulses using a computer. 
We used control pulses of the form
\begin{equation}
\Omega(0,t)=\Omega_0[\mathrm{sech}(t/\tau)]^{1+i\mu} 
\end{equation}
which yield a time dependent detuning from the atomic line center as:
\begin{equation}
\partial_t\Phi(0,t)=-\frac{\mu}{\tau}\mathrm{tanh}\left(\frac{t}{\tau}\right).
\end{equation}
For $\mu=0$, there is no chirp and the pulse area is 
$\mathcal{A}=\pi\Omega_0/\tau$, while for $\mu\neq0$ the chirp ranges from $\mu/\tau$
to $-\mu/\tau$.
Before the first control pulse, all atoms are in the ground state,
while the second one propagates through the medium prepared by the first one -
atomic excitations remain, but the coherences have had time to dephase.
Having obtained $\Omega_j(z,t)$ we constructed the operators $\hat{U}^{Cj}(z,t)$ to
investigate its matrix elements as  a function of $\Delta$ and $z$.
We considered two different cases. In one, the inhomogeneous broadening of the atomic line is
much larger than the pulse bandwidth, so $g(\Delta)=g_0$ is taken to be constant. In this case the
pulses are able to invert only a part of the ensemble,
leaving atoms with a large $\Delta$ untouched.
Clearly, there is then a transition region where the control pulses interact with the atoms but AP
is not perfect.
In the other case, we have a Gaussian line shape function 
$g(\Delta)=\exp(-\Delta^2/2\sigma_\Delta^2)/\sigma_\Delta\sqrt{2\pi}$ and control pulse bandwidth
is great enough to encompass the whole absorption line.
The first of these two cases is especially interesting, as it is the one that corresponds to the
case of a very widely broadened ionic transition in a crystal.

\begin{figure}[htb]
\includegraphics[width=0.48\textwidth]{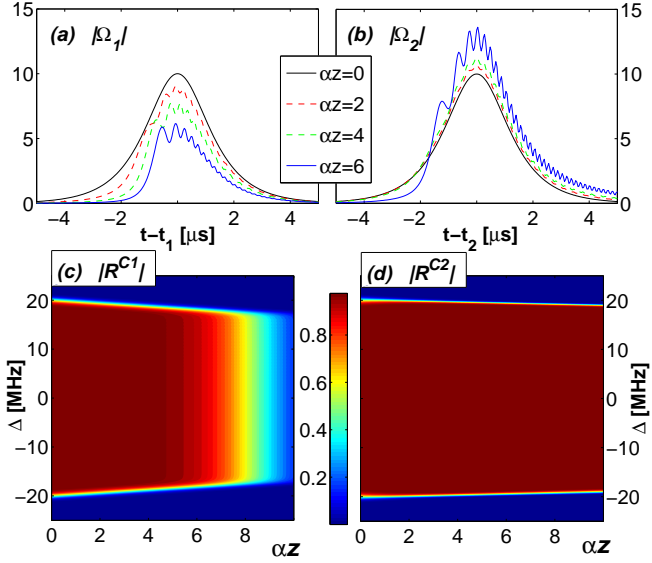}
\caption{The behavior of two successive frequency-chirped sech control pulses as they propagate
in the storage medium with $g(\Delta)=g_0$. Pulse parameters: $\tau=1 \mathrm{~\mu s} $,
$\Omega_0=10 \mathrm{~MHz}$ and
$\mu=-20$. (a) and (b): $|\Omega_1(t)|$
and
$|\Omega_2(t)|$ (both in MHz) at normalized propagation distances of $\alpha_dz=0,2,4,6$. (c) and
(d): contour plots of $|\mathcal{R}^{C1}|$ and  $|\mathcal{R}^{C2}|$ vs. $\Delta$
and $\alpha_dz$.}
\label{fig_pulses}       
\end{figure}

\begin{figure}
\includegraphics[width=0.48\textwidth]{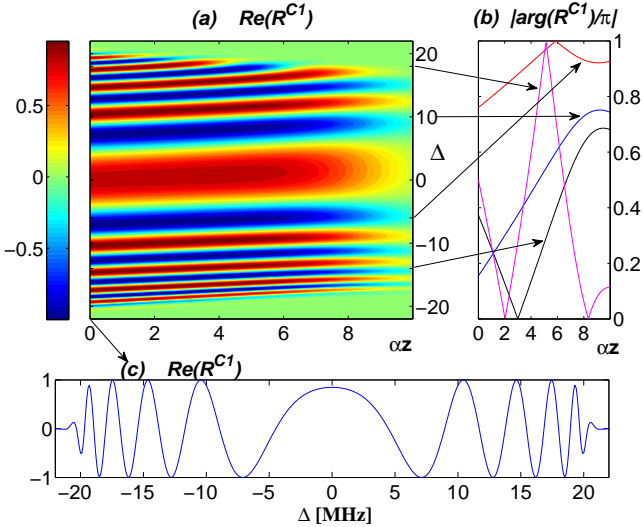}
\caption{The behavior of $\mathcal{R}^{C1}$ for the first control pulse. 
(a) Contour plot of 
$Re(\mathcal{R}^{C1})$ as a function of $\Delta$ (MHz) and $\alpha_dz$. 
(b) Line plots of $|arg(\mathcal{R}^{C1})/\pi|$ vs. ${\alpha_d}z$ for
$\Delta=-14$, $-6$, $10$ and $18$, indicated by arrows.
(c) Line plot of $Re(\mathcal{R}^{C1})$ vs. $\Delta$ at ${\alpha}_dz=0$.
(Pulse parameters are the same as for Fig. \ref{fig_pulses})}
\label{fig_RC1}       
\end{figure}

Figure \ref{fig_pulses} (a) and (b) depict how a pair of successive chirped control
pulses are deformed during propagation when $g(\Delta)=g_0$. The time plots of the pulse amplitudes
show
clearly that the first pulse is considerably attenuated, while the second one is amplified.
At the same
time both pulse amplitudes are modulated in time. Figs. \ref{fig_pulses} (c) and (d) depict  
$|\mathcal{R}^{C1}|$ and  $|\mathcal{R}^{C2}|$ as a function of $\Delta$ and
$\alpha_dz$. They show that at $z=0$ both pulses create AP over roughly the 
$\Delta\in\{-20\mathrm{~MHz},20\mathrm{~MHz}\}$ frequency interval, but the range where AP works for
the first
pulse narrows continuously, and at about $\alpha_dz=4.5$ it starts deteriorating over the entire
frequency range. The second pulse on the other hand maintains AP until the calculated distance
of $\alpha_dz=10$ with only the frequency interval narrowing very slightly.
Figure \ref{fig_RC1} illustrates the rephasing power of the first control pulse, or rather the lack
of it. The contour plot of
$Re(\mathcal{R}^{C1})$ [Fig.\ref{fig_RC1} (a)] shows that the phase associated with the
transformation of the atomic coherences is not uniform across the ensemble, not even in the domain
where the pulse creates AP. It changes with
$\Delta$ at any given optical depth $\alpha_dz$ and also for any $\Delta$ as a function of
the optical depth $\alpha_dz$. Line plots of $|arg(\mathcal{R}^{C1})/\pi|$ for several values of
$\Delta$ in Fig.\ref{fig_RC1} (b) and of $Re(\mathcal{R}^{C1})$ at $\alpha_dz=0$ in
Fig.\ref{fig_RC1} (c) demonstrate this even more clearly. 

\begin{figure}
\includegraphics[width=0.48\textwidth]{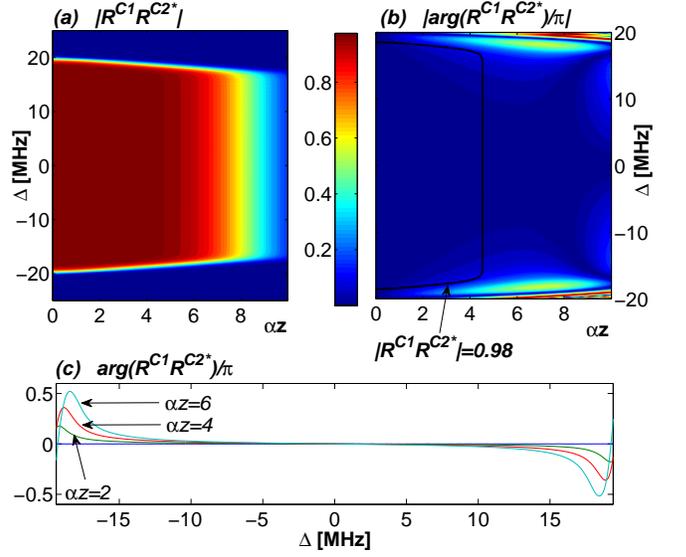}
\caption{The behavior of $\mathcal{R}^{C1}[\mathcal{R}^{C2}]^*$ for a pair of successive
chirped control pulses. 
(a) Contour plot of $|\mathcal{R}^{C1}[\mathcal{R}^{C2}]^*|$ as a function of $\Delta$ and
$\alpha_dz$.
(b) Contour plot of $|arg(\mathcal{R}^{C1}[\mathcal{R}^{C2}]^*)/\pi|$ - the heavy black line
corresponds to $|\mathcal{R}^{C1}[\mathcal{R}^{C2}]^*|=0.98$. 
(c) Line plots of $arg(\mathcal{R}^{C1}[\mathcal{R}^{C2}]^*)/\pi$ vs. $\Delta$ at optical depths of
$\alpha_dz=0,2,4, \mathrm{~and~} 6$.
(Pulse parameters are the same as for Fig. \ref{fig_pulses}).}
\label{fig_RC1RC2}       
\end{figure}

What we have seen so far is just what we anticipated. The surprising result is shown
in Fig. \ref{fig_RC1RC2} where the behavior of $\mathcal{R}^{C1}[\mathcal{R}^{C2}]^*$ has been
plotted, the quantity associated with coherence
rephasing by a pair of two successive control pulses.
Its magnitude, shown in Fig. \ref{fig_RC1RC2} (a) gives the probability that an atom of the
ensemble at $z$ and with frequency offset $\Delta$ undergoes AP twice as a result of the
interaction. This value is close to one in
an extended region of $\Delta$ and $\alpha_dz$ - a region essentially identical to the one in
which the first control pulse is able to create AP [see Fig. \ref{fig_RC1} (c)]. 
Remarkably, the complex phase $arg(\mathcal{R}^{C1}[\mathcal{R}^{C2}]^*)$
shown in Fig. \ref{fig_RC1RC2} (b)
is also essentially constant in this region.
The line where $|\mathcal{R}^{C1}[\mathcal{R}^{C2}]^*|=0.98$ has been drawn over the contour plot
for guidance. This means that despite the considerable and unequal distortion that the two control
pulses suffer during propagation, the pair of chirped pulses can rephase a sizable domain of the
atomic ensemble both in terms of optical depth and frequency interval. With these parameters
the boundaries are roughly at $\Delta\in\{-15\mathrm{~MHz},15\mathrm{~MHz}\}$ and $\alpha_dz=4.5$,
but this can be extended easily by increasing the pulse amplitude or the chirp slightly. For
example, the same pulses with $\Omega_0=12\mathrm{~MHz}$ instead of $\Omega_0=10\mathrm{~MHz}$ can
rephase the coherences to
about $\alpha_dz=8.7$.

\begin{figure}
\includegraphics[width=0.48\textwidth]{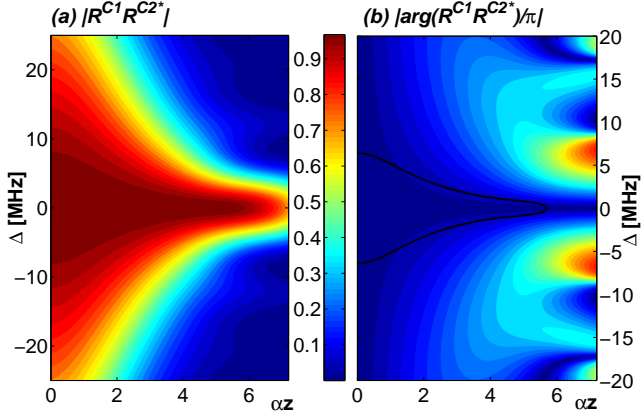}
\caption{The behavior of $\mathcal{R}^{C1}[\mathcal{R}^{C2}]^*$ for a pair of successive $\pi$
pulses. 
(a) Contour plot of $|\mathcal{R}^{C1}[\mathcal{R}^{C2}]^*|$ as a function of $\Delta$ and
$\alpha_dz$.
(b) Contour plot of $|arg(\mathcal{R}^{C1}[\mathcal{R}^{C2}]^*)/\pi|$ - the heavy black line
corresponds to $|\mathcal{R}^{C1}[\mathcal{R}^{C2}]^*|=0.98$.
Pulse parameters: $\tau=0.01 \mathrm{~{\mu}s}$, $\Omega_0=100 \mathrm{~MHz}$, $\mu=0$.}
\label{fig_pipulse}       
\end{figure}

For a comparison, we also calculated the rephasing abilities of a pair of consecutive $\pi$-pulses
in an identical way.
Naturally, a pulse of much shorter duration and hence much greater peak intensity is needed to
rephase a
comparable region of the ensemble. Figure \ref{fig_pipulse} shows the contour plots 
of the magnitude and phase of $\mathcal{R}^{C1}[\mathcal{R}^{C2}]^*$. 
It is clear, that with the chosen parameters ($\tau=0.01 \mathrm{~{\mu}s}$, $\Omega_0=100
\mathrm{~MHz}$, $\mu=0$) the performance of the $\pi$-pulse pair is inferior to that
of the chirped pulse pair.
The frequency interval where $|\mathcal{R}^{C1}[\mathcal{R}^{C2}]^*|\approx 1$
is much narrower even at $z=0$ and narrows rapidly. While the pulse energies are the same with
these parameters, the peak intensity of the $\pi$-pulses is 100 times greater. 

One advantage of $\pi$-pulses is of course that
the interaction time is much shorter, the control works faster. However, because
of the long 'tail' that the $\pi$-pulses develop during propagation \cite{Ruggiero2009} this
advantage is far smaller than the actual difference between the time constants. 
(For the present case the initial $\pi$
pulses of $\tau=0.01 \mathrm{~\mu s}$ widen to several times $0.1 \mathrm{~\mu s}$ by about
$\alpha_dz=5$ which means that an initial advantage of two orders of magnitude essentially reduces
to one order of magnitude.) 

\begin{figure}
\includegraphics[width=0.48\textwidth]{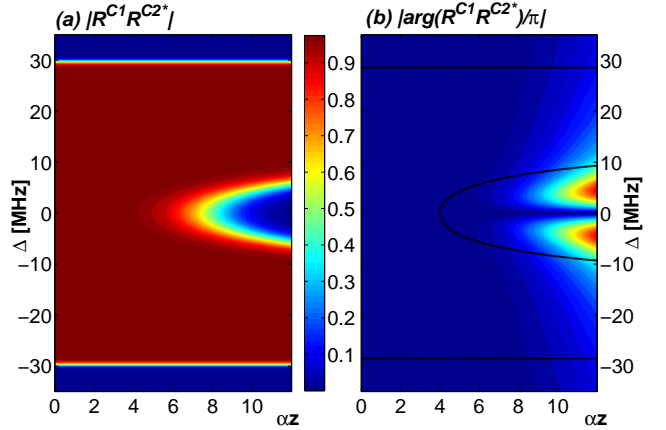}
\caption{The behavior of $\mathcal{R}^{C1}[\mathcal{R}^{C2}]^*$ for a pair of successive
frequency-chirped
 pulses propagating through a medium with a Gaussian inhomogeneous line shape function
with $\sigma_\Delta=6.2666$. 
(a) Contour plot of $|\mathcal{R}^{C1}[\mathcal{R}^{C2}]^*|$ as a function of $\Delta$ and
$\alpha_dz$.
(b) Contour plot of $|arg(\mathcal{R}^{C1}[\mathcal{R}^{C2}]^*)/\pi|$ - the heavy black line
corresponds to $|\mathcal{R}^{C1}[\mathcal{R}^{C2}]^*|=0.98$.
Pulse parameters: $\tau=1 \mathrm{~{\mu}s}$, $\Omega_0=12 \mathrm{~MHz}$, $\mu=-30$.}
\label{fig_gaussiang}       
\end{figure}

Finally, Fig. \ref{fig_gaussiang} depicts $\mathcal{R}^{C1}[\mathcal{R}^{C2}]^*$ for a pair of
chirped control pulses that propagate through a medium with a relatively narrow inhomogeneous
broadening. $g(\Delta)$ is now a Gaussian with a width of $\sigma_\Delta=6.2666$, while
the chirp range of the pulses is 
from -30 MHz to 30 MHz, so the control pulses are able to invert the whole atomic ensemble.
Fig. \ref{fig_gaussiang} (a) shows that now the ability of the pulse pair to create AP twice
is lost only around the central frequencies where the medium is optically the densest. 
Fig. \ref{fig_gaussiang} (b) shows that again, $arg(\mathcal{R}^{C1}[\mathcal{R}^{C2}]^*)$ is
almost constant in the region where AP works (the black line again marking the boundary of
$|\mathcal{R}^{C1}[\mathcal{R}^{C2}]^*|=0.98$, a deviation from the constant phase can be observed
for $\alpha_dz>7$).

\section{Photon echos}

To verify that frequency-chirped control pulses are suitable for applications in photon-echo
memories, we used Eqs. \ref{eq_bloch1}, \ref{eq_maxwell1} and \ref{eq_polarization} to calculate
the echos of a set of weak signal pulses and compare them with the original signal. Gaussians of the
form $E_s(t)\sim exp(-t^2/2\tau^2)$ were used with $\tau=1\mathrm{~\mu s}$, and a
variable frequency $\omega_s$, detuned slightly from 
$\omega_0$ (to which the central frequency of the control pulses was tuned). We performed a
parameter scan with respect to $\omega_s$ and  the optical length of the storage medium $\alpha_dL$
 for both
variants of the memory protocol described in subsection \ref{sec_protocols}. We used a classical
signal, but
assumed that it is so weak that it does not in any way influence the propagation of the strong
control pulses. Thus after having calculated the coherences imprinted in the ensemble by the signal,
we used the time evolution operators computed in Sec. \ref{sec_rephasing} (without a signal) 
to calculate the atomic states at $t_3-T$. 
We then solved Eqs. \ref{eq_bloch1},
\ref{eq_maxwell1} and \ref{eq_polarization} again numerically for the time interval $[t_3-T,t_3+T]$
to obtain the echo.

The efficiency of the memory protocol with chirped pulses was then characterized by calculating
the ratio of echo energy to signal energy
\begin{equation}
\eta=\frac{\int|E_{e}(t)|^2 dt}
{\int|E_{s}(t)|^2 dt}
\end{equation}
which, in the weak signal limit corresponds to the overall probability that an incident photon is
absorbed by the medium and later 
re-emitted as a signal echo. Another figure of merit calculated was a classical fidelity
\begin{equation}
\xi=\max_{\substack{t_{delay}}}\left|\frac{\int E_{e}(t-t_{delay})E_s^*(t) dt}
{\sqrt{\int|E_{s}(t)|^2 dt\times\int|E_{e}(t)|^2 dt}}\right|
\end{equation}
which characterizes the similarity of the signal and echo fields, neglecting an arbitrary difference
in phase and reduction in amplitude. 

Our calculation of the echo field includes all of the atomic ensemble, those atoms that
undergo AP twice during the interaction with the control pulses, and also those that do not. Atoms
that are too far either in optical depth $\alpha_dz$ or in frequency offset $\Delta$ to be
rephased, may still contribute during echo emission, possibly to distort the signal.
However, the calculation is entirely classical, so it does not account for quantum noise, such as
spontaneously emitted photons from atoms that, due to imperfect AP, are left in the excited
state after the second control pulse. The classical fidelity presented here cannot be
identified with the true fidelity of a one (few) photon signal pulse.

\begin{figure}
\includegraphics[width=0.48\textwidth]{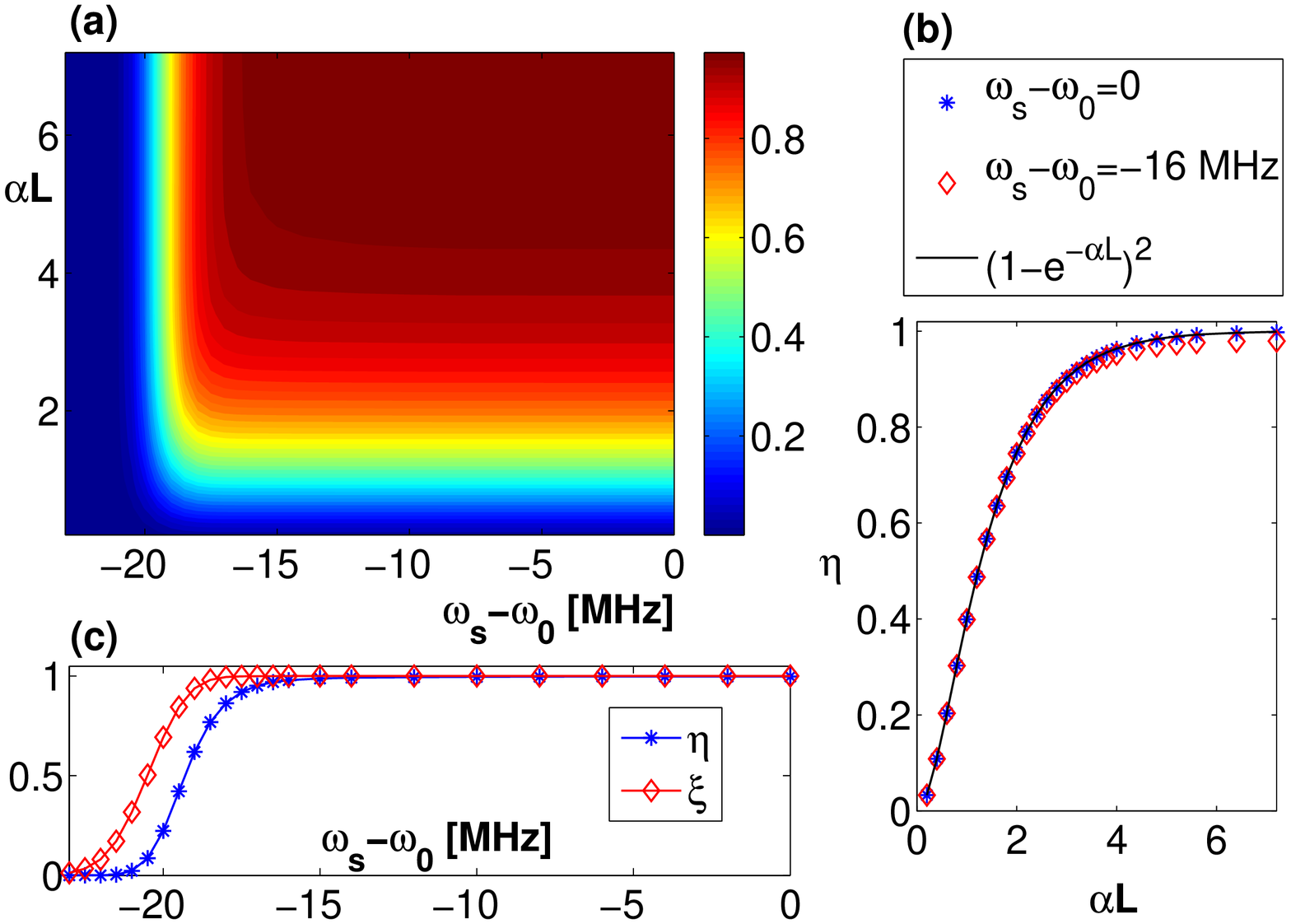}
\caption{Memory efficiency and fidelity for backward echo emission and $g(\Delta)=g_0$. 
(a) $\eta$ vs. optical length ${\alpha_d}L$ and signal detuning
$\omega_s-\omega_0$. (b) $\eta$ vs. ${\alpha_d}L$ for $\omega_s-\omega_0=0$ (blue $\ast$), 
$\omega_s-\omega_0=-16\mathrm{~MHz}$ (red $\diamond$) and $\eta'=(1-e^{-\alpha_dL})^2$ (solid
line). (c) $\eta$ and $\xi$ vs. $\omega_s-\omega_0$ at
${\alpha_d}L=7.2$. Control pulse parameters are identical to those in Figs. \ref{fig_pulses},
\ref{fig_RC1} and \ref{fig_RC1RC2}.
}
\label{fig_eff1}       
\end{figure}

\begin{figure}
\includegraphics[width=0.48\textwidth]{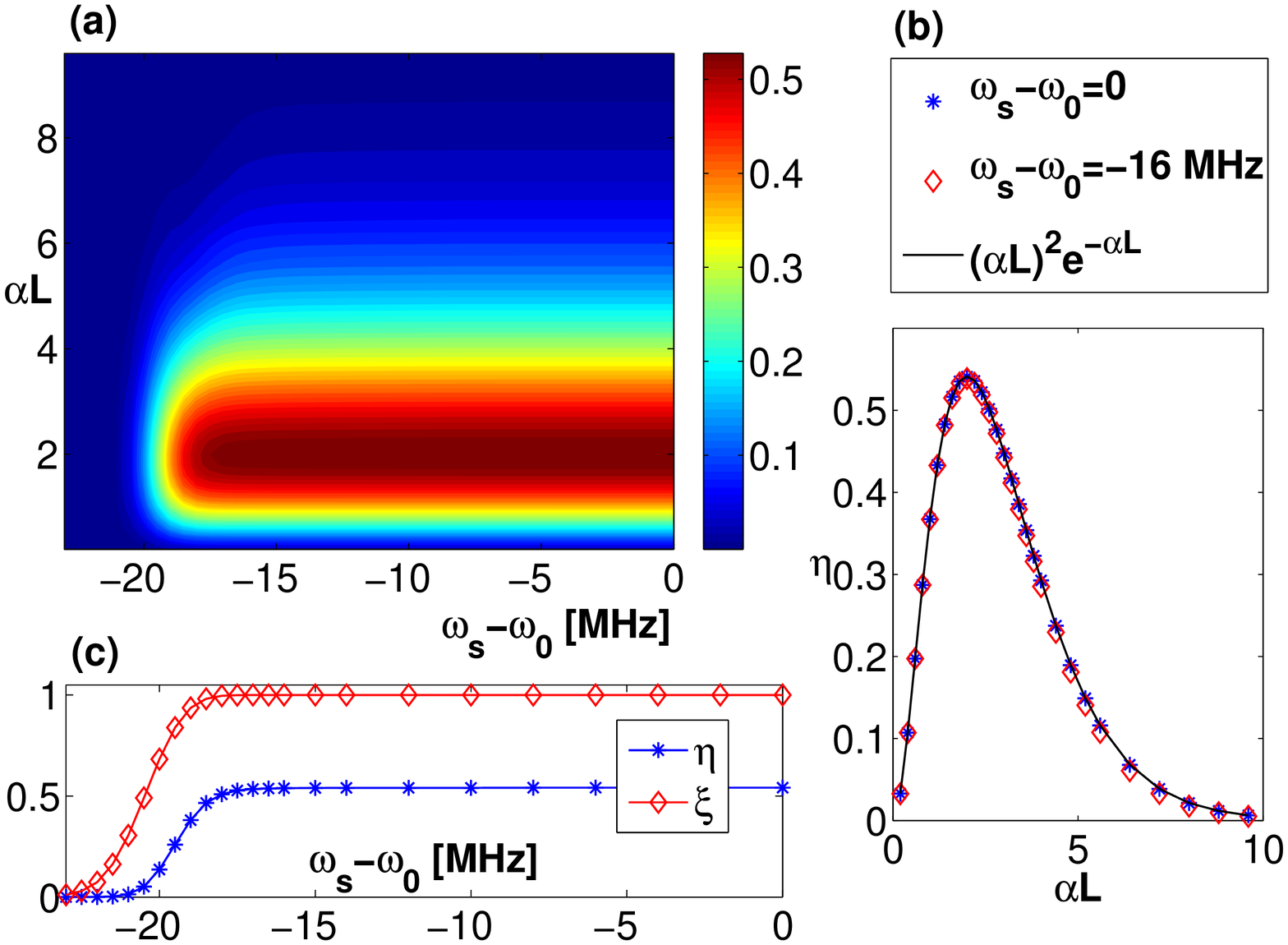}
\caption{Memory efficiency and fidelity for forward echo emission and $g(\Delta)=g_0$. 
(a) $\eta$ vs. optical length ${\alpha_d}L$ and signal detuning
$\omega_s-\omega_0$. 
(b) $\eta$ vs. ${\alpha_d}L$ for $\omega_s-\omega_0=0$ (blue $\ast$), 
$\omega_s-\omega_0=-16\mathrm{~MHz}$ (red $\diamond$) and
$\eta'=(\alpha_dL)^2e^{-\alpha_dL}$ (solid
line).
(c) $\eta$ and $\xi$ vs. $\omega_s-\omega_0$ at
${\alpha_d}L=2$. Control pulse parameters are identical to those
in Figs. \ref{fig_pulses}, \ref{fig_RC1} and \ref{fig_RC1RC2}.}
\label{fig_eff2}       
\end{figure}

\begin{figure}
\includegraphics[width=0.48\textwidth]{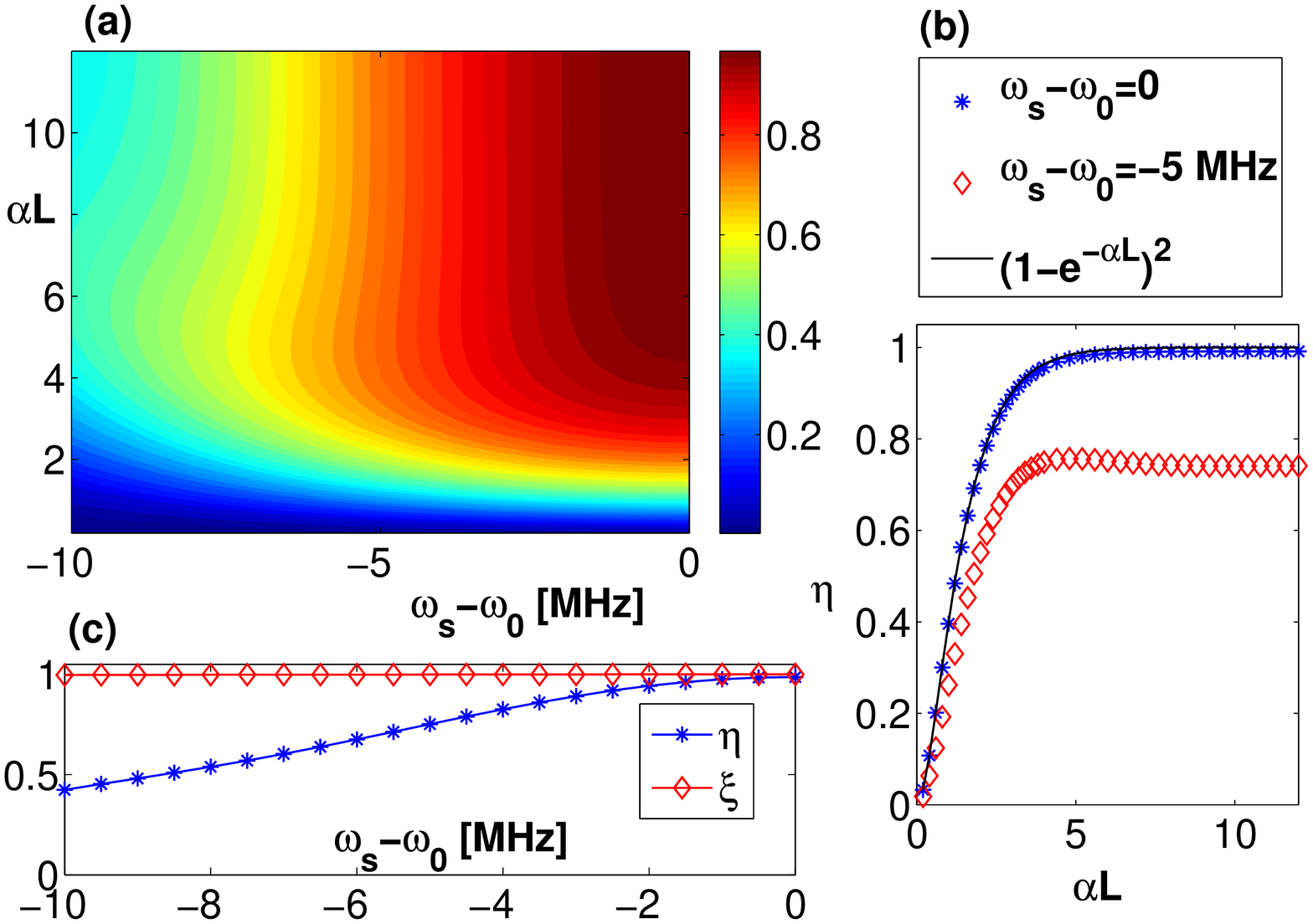}
\caption{Memory efficiency and fidelity for backward echo emission and $g(\Delta)$ a
Gaussian. (a) $\eta$ vs. optical length ${\alpha_d}L$ and signal detuning
$\omega_s-\omega_0$.
(b) $\eta$ vs. ${\alpha_d}L$ for $\omega_s-\omega_0=0$ (blue $\ast$), 
$\omega_s-\omega_0=-5\mathrm{~MHz}$ (red $\diamond$) and $\eta'=(1-e^{-\alpha_dL})^2$ (solid
line). (c) $\eta$ and $\xi$ vs. $\omega_s-\omega_0$ at
${\alpha_d}L=6$. Control pulse parameters are identical to those 
in Fig. \ref{fig_gaussiang}.}
\label{fig_eff3}       
\end{figure}

\begin{figure}
\includegraphics[width=0.48\textwidth]{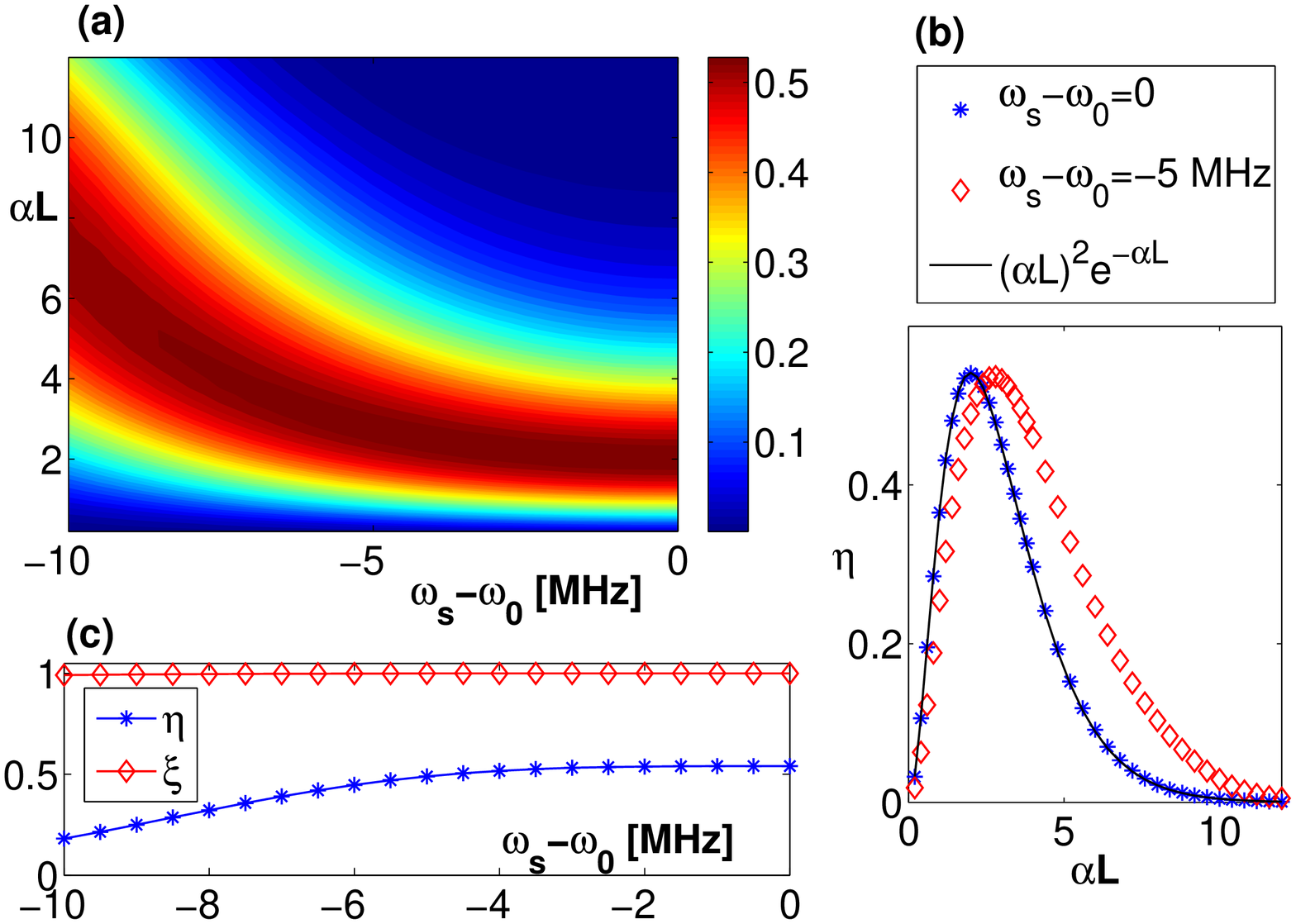}
\caption{Memory efficiency and fidelity for forward echo emission and $g(\Delta)$ a
Gaussian. 
(a) $\eta$ vs. optical length ${\alpha_d}L$ and signal detuning
$\omega_s-\omega_0$. (b) $\eta$ vs. ${\alpha_d}L$ for $\omega_s-\omega_0=0$ (blue $\ast$), 
$\omega_s-\omega_0=-5\mathrm{~MHz}$ (red $\diamond$) and
$\eta'=(\alpha_dL)^2e^{-\alpha_dL}$ (solid
line). (c) $\eta$ and $\xi$ vs. $\omega_s-\omega_0$ at
${\alpha_d}L=2$. Control pulse parameters are identical to those 
in Fig. \ref{fig_gaussiang}.}
\label{fig_eff4}       
\end{figure}

Figures \ref{fig_eff1}-\ref{fig_eff4} depict our results. In each figure, (a)
shows a contour plot of the efficiency $\eta$ as a function of signal detuning $\omega_s-\omega_0$
and optical length $\alpha_dL$. The results are symmetric with respect to
$\omega_s-\omega_0$, so only negative values have been plotted for a better visibility. (b) in each
figure shows $\eta$ for two specific values of $\omega_s-\omega_0$ along with the curves of the
best theoretical efficiency calculated for CRIB \cite{Sangouard2007}: $\eta'=(1-e^{\alpha_dL})^2$
for backward echo emission and
$\eta'=(\alpha_dL)^2e^{-\alpha_dL}$ for forward echo emission. In each figure, (c)
is a plot of $\eta$ and the fidelity $\xi$ as a function of $\omega_s-\omega_0$ for a given optical
length. 
Figures \ref{fig_eff1} and \ref{fig_eff2} show that when we have $g(\Delta)=g_0$, $\eta$ is
practically constant at any $\alpha_dL$ for a wide range of signal detunings. The efficiencies for
$\omega_s=\omega_0$ lie precisely on the curves for $\eta'$, 
and even $\omega_s-\omega_0=-16\mathrm{~MHz}$ yields efficiencies just very slightly below them.
The fidelity $\xi$ is also extremely close to unity in this region. Both $\eta$ and $\xi$ start to
decrease
only when the boundary of the control pulse frequency range (between -20 MHz to 20
MHz) is approached. 
For backward echos, $\eta_{max}=1$ with $\eta=0.9985$ being reached by $\alpha_dL=7.2$ depicted in
Fig. \ref{fig_eff1} (b) and (c), while for forward echos $\eta_{max}=0.54$ at $\alpha_dL=2$
[Fig. \ref{fig_eff2} (b) and (c)].
The reason for the reduction of $\eta_{max}$ for forward echos is 
the same as in the case of CRIB - the echo
is reabsorbed again by the storage medium if it is too thick. 

Figures \ref{fig_eff3} and \ref{fig_eff4} show the case when we have a Gaussian
$g(\Delta)$. Backward echo efficiency now approaches $\eta_{max}=1$ only for 
$\omega_s=\omega_0$ and is considerably less for a signal detuning of
$\omega_s-\omega_0=-5\mathrm{~MHz}$ already [Figs. \ref{fig_eff3} (a) and (b)].
Forward echo efficiency does approach $\eta_{max}=0.54$ for a wider range of detunings, but the
storage medium length which is required is greater for a signal detuned from $\omega_0$ 
[Figs. \ref{fig_eff4} (a) and (b)]. This is because with a relatively narrow broadening
($\sigma_\Delta=6.2666$) a signal detuned from the atomic line center experiences a reduced optical
length. For the same reason, at a given medium length, the efficiency $\eta$ decreases with
$|\omega_s-\omega_0|$. 
At the same time Figs. \ref{fig_eff3} and \ref{fig_eff4} (c) show that the 
the echo signal is only reduced but not distorted, $\xi$ remains close to one.

Some comments on possible control pulse parameters are in order. First of all, decoherence
effects other than dephasing due to inhomogeneous broadening have been neglected in our
description. Because there are no other 'inherent' timescales, the results presented apply equaly
well to any other parameter set which is scaled consistently (pulse lengths, Rabi frequencies,
chirp parameters and atomic frequency offsets must be scaled together). 
Naturally, the time delay between the control pulses (which is half the memory
storage time without the auxiliary shelving state $|s\rangle$) cannot be more than a few percent of
the excited state lifetime $\Gamma^{-1}$ for our results to remain valid. While control pulses of
arbitrary $\tau$ can satisfy the requirements of AP, the upper bound will be set by
$\tau^{max}\sim 10^{-3}\Gamma^{-1}$. AP requires that the pulse
area of the real envelope function be $\mathcal{A}=\int|\Omega(t)|dt\gtrsim 10 \pi$. 
(The precise value
of course depends slightly on the pulse shape and the chirp function, in our case
$\mathcal{A}=10\pi$ proved entirely sufficient.)
Together with the constraint on $\tau^{max}$, this sets the lower bound of the peak Rabi
frequency to be $\Omega_0^{min}\gtrsim10^4 \Gamma$.  

The ideal choice for the chirp parameter is such that the full bandwidth of the pulse is at least a
few times $\Omega_0$. Much lower values are impractical, because then the extension of the
transition region in $\Delta$ where the control pulses perturb the atoms but AP fails is comparable
to the region where AP works correctly. (The significance of this will be clarified shortly.) 
$\tau^{min}$ and $\mu^{max}$ will be set by the
requirement that the full control pulse bandwidth cannot exceed the distance to the nearest unused
electronic level. This then constrains $\Omega_0^{max}$ as well. It may well happen, however, that
this $\Omega_0^{max}$ already corresponds to a peak intensity that is either too high to generate or
for the medium (host crystal) to endure. In this case the latter constraint on $\Omega_0^{max}$ 
obtains precedence and sets $\tau^{min}$ via the requirement on $\mathcal{A}$. 

Regarding the possible range of optical depths one may consider, we note that due to the exponential
decay of the signal during absorption, a medium with $\alpha_dL=5-10$ is perfectly sufficient to
absorb the signal. Thus we need not consider optical depths of $\alpha_dz>10$ and, indeed the ideal
choice for forward echo emission is $\alpha_dL=2$. 
In our simulations $\Omega_0=10\mathrm{~MHz}$ rephases atoms of the ensemble almost perfectly until
about $\alpha_dz=4.5$, which is already sufficient for an excellent memory efficiency.
However, this limit can easily be extended if necessary - $\Omega_0=12\mathrm{~MHZ}$ rephases the
ensemble to $\alpha_dz=8.7$, $\Omega_0=14\mathrm{~MHZ}$, to well above $\alpha_dz=10$.

\subsection{Further implications for photon-echo memories}

In light of these results, it is clear that a pair of chirped control pulses that 
drive AP are better for building quantum memories than a pair of $\pi$-pulses
for several reasons. The first one mentioned already in the introduction is of
course that pulses with much smaller peak intensity can be used. 
(In Sec. \ref{sec_rephasing} the particular example showed that chirped pulses with two orders of
magnitude smaller peak intensity deliver far better rephasing ability for the case of constant
$g(\Delta)$.) 

The second advantage is the small width of the transition region in frequency where
atoms are not perfectly rephased, but nevertheless considerably perturbed by the control pulses.
When we have a widely broadened inhomogeneous line and can only hope to rephase a relatively narrow
frequency region (which is in fact the generic case in rare-earth doped crystals), this is very
important because atoms that are not inverted twice perfectly may remain excited after the second
control pulse. They will then be a source of noise due to spontaneous emission at the time of signal
retrieval. (It is important to note that even though the duration $T_{mem}$ of the whole sequence
may
be $T_{mem}\ll\Gamma^{-1}$, the question of spontaneous emission during echo emission must still
be considered, at least qualitatively. Because the ultimate goal is to retrieve a single photon
pulse, if there are a large number of excited atoms in the ensemble at retrieval time, spontaneous
emission may be detrimental however short the signal pulse is \cite{Ruggiero2009}.)

\begin{figure}
\includegraphics[width=0.48\textwidth]{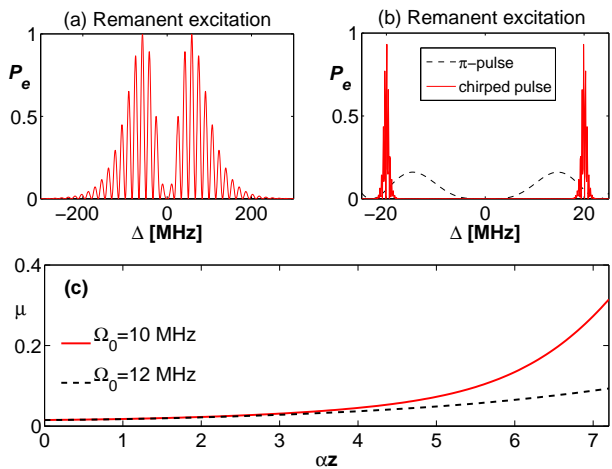}
\caption{Remanent excitation in the medium after the two control pulses at ${\alpha_d}z=0$ for $\pi$
pulses [(a) and dashed line in (b)] and chirped pulses [solid line in (b)] as function of
$\Delta$. Pulse parameters are the same as for Figs \ref{fig_pulses} and \ref{fig_pipulse}. (c) The
ratio $\mu$ of the overall excitation left in the medium by the chirped control pulses and the $\pi$
control pulses as a function of optical depth for two different chirped pulse amplitudes.}
\label{fig_excitation}       
\end{figure}

To assess the reduction in spontaneous noise more quantitatively, we compute the probabilities that
atoms remain in the excited state after the second - $\pi$ or chirped - control pulse, the 'remanent
excitation' due to the control pulses, $P_e^\pi(\Delta,z)$ and $P_e^C(\Delta,z)$. This quantity can
be extracted from the
time evolution operator used earlier:
\begin{equation*}
 P_e(\Delta,z)=\left|\left[\hat{U}(\Delta,z)\right]_{2,1}\right|^2.
\end{equation*}
The plot of $P_e^\pi(\Delta)$ at $\alpha_dz=0$ for a
pair of $\pi$-pulses can be seen in Fig. \ref{fig_excitation} (a), which shows two wide regions of
remanent excitation on both sides of the
narrow central hole. This latter is the region where atoms are correctly rephased, while the fast
oscillations on both sides trace out a slow envelope of two wide maxima. The precise frequency and
phase of the rapid oscillations depends on the time between the two control pulses, the width of
the slow envelope however only depends on their bandwidth. In this case,
$\tau=0.01\mathrm{~{\mu}s}$ has been used 
(as for Fig. \ref{fig_pipulse}), so the width of the high $P_e^\pi$ region is about 100 MHz.
$P_e^C(\Delta)$ is
shown by the solid line in Fig. \ref{fig_excitation} (b), which again shows a rapidly oscillating
curve that traces out two relatively narrow maxima at $\Delta=\pm20\mathrm{~MHz}$. (Pulse parameters
used were the same as for Figs. \ref{fig_pulses}, \ref{fig_RC1} and \ref{fig_RC1RC2}, $\tau=1
\mathrm{~\mu s} $,
$\Omega_0=10 \mathrm{~MHz}$ and $\mu=-20$). The width of the maxima is approximately 1 MHz, the
bandwidth of the pulse due solely to its duration, while their positions are at the two
limits of the full frequency range of the chirped pulse. The dashed curve between the two sharp
maxima is the plot of the very central part of $P_e^\pi(\Delta)$ from Fig. \ref{fig_excitation} (a),
plotted to show that with these parameters, the $\pi$-pulse pair rephases atoms only in a much
smaller frequency range.

To characterize the reduction in spontaneous noise due to atoms in the transition
region, we calculate
$\mu=\int P_e^C(\Delta)d\Delta/\int P_e^\pi(\Delta)d\Delta$, the ratio of the overall
excitation remaining in the two cases. This quantity is shown with a solid line in Fig.
\ref{fig_excitation} (c), its initial value at $\alpha_dz=0$ is $\mu=0.015$.
It increases very slightly at first with the optical depth, because as the $\pi$ pulse loses energy 
its bandwidth decreases and thus the width of the transition region narrows somewhat.
At around the point where AP for the first chirped control pulse starts failing (${\alpha_d}z=4.5$
in this case)
$\mu$ starts increasing much faster because the second chirped pulse then starts leaving more and
more atoms in the excited state for every $\Delta$ within its bandwidth. At ${\alpha_d}z=4.5$ we
have $\mu=0.056$, so spontaneous noise due to remanent excitation at this point is still about 18
times less for the chirped pulses. 
The optical depth to which AP works can also
be extended easily with slightly larger pulse amplitudes - $\mu$ calculated with a chirped pulse
amplitude of $\Omega_0=12\mathrm{~MHz}$ is shown with a broken line in Fig. \ref{fig_excitation}
(c). We also note that for the comparison we used $\pi$-pulses which are capable of rephasing
a far smaller frequency domain within the ensemble to start with. (The width of the
central region where atoms are rephased is only about 12 MHz, while it is close to 34 MHz for the
chirped pulses, see Fig. \ref{fig_excitation} (b). Also compare Figs. \ref{fig_RC1RC2} and
\ref{fig_pipulse}.) 
Altogether it is safe to say that spontaneous
emission induced noise can be reduced by a factor of $10^{-2}$ if we use 
chirped control pulses instead of $\pi$-pulses
with comparable rephasing ability.

For a relatively narrow $g(\Delta)$ it is of course possible to choose control pulses where the
whole atomic ensemble is inverted, i.e. the transition region  lies in a spectral domain void of
 absorbers. Then its width is not important. 
However, unmanipulated ionic transition lines in solids are usually
broad. A narrow absorption feature (of a few MHz
in width as used in Sec. \ref{sec_rephasing}) would have to be prepared using techniques
identical to those used for CRIB.

One may also envision devices where the same storage medium is used
for several distinct 'memory channels' of different frequency, manipulated separately by control
pulses. In this case, one clearly has to maintain a spectral distance between
the channels such that they do not interfere with each other - 'crosstalk' between the channels must
be kept low. This means that the interchannel spectral distance is constrained by the width of the
transition region. Chirped pulses clearly have a much greater potential in this field. 

\section{Summary and Conclusion}

In this paper, we have investigated the ability of a pair of chirped control pulses to rephase the
coherences in an inhomogeneously broadened, optically thick ensemble of two-level atoms. By solving
the Maxwell-Bloch equations numerically, we have shown that as long as both pulses drive
AP between the atomic states, they can rephase collectively the atomic
coherences. This result is somewhat counterintuitive, because the 
time integral of the adiabatic eigenvalues plays an important role in rephasing and the two
successive pulses evolve differently as they propagate through the medium. The first pulse is
attenuated
because of absorption, while the second one, propagating in the gain medium prepared by the first
one, is amplified. Nevertheless, there is a well defined region in the ensemble in terms of atomic
frequency and optical depth, where rephasing works well. 
The extent of this region is considerably greater than that rephased by a pair of consecutive $\pi$
pulses with the same energy, but two orders of magnitude higher peak intensity, which
is an important
property when rare-earth ion impurities embedded is a crystal are used as a storage medium.
The price to pay is a somewhat longer control pulse time, which, however is only about one order of
magnitude greater after pulse propagation effects are taken into account.

We have shown that it is possible to use chirped control pulses in photon-echo memory schemes,  
where the primary echo after the first control pulse is silenced by spatial phase mismatching.
The atomic coherences rephase again after the second control pulse, this time without the storage
medium being inverted. Using chirped control pulses, the same maximum echo efficiencies are
theoretically attainable in an unmanipulated, 'naturally' inhomogeneously broadened ensemble,
as in schemes such as CRIB or AFC. For these latter schemes numerous preparatory steps are
required to obtain the absorption feature required by the protocol.
For chirped control pulses, the frequency width of the transition region where the control
pulses excite the atoms considerably, but fail to rephase them properly can be relatively small.
This means, that quantum noise emanating from it (atoms left in their excited
state by the control pulses emitting photons spontaneously during echo emission) could be small
enough for the retrieval
of quantum information. 

It is also possible to use an ensemble with great inhomogeneous width 
for the storage of several memory channels with different frequencies simultaneously.
Because of the narrow transition region, using chirped pulses
means that a much smaller frequency distance between the distinct channels is needed to suppress
crosstalk between them. 
This allows multimode information storage 
with the separate, on demand recall of the information stored in different channels.
The same with a pair of $\pi$-pulses would not be possible, for the width of the disturbing
transition region is orders of magnitude greater.

\appendix*
\section{Construction of the time evolution operator}

Let us regard an atom at $z$ and with frequency offset $\Delta$, such that it is well within the
frequency range spanned by the spectrum of the control pulse.
$\hat{U}{(\Delta,z)}$ propagates the probability amplitudes from $t_0+T$ to $t_3-T$ as
\begin{equation*}
\left(\begin{array}{c}\alpha_n'\\ \beta_{n\pm1}'\end{array}\right)=
\hat{U}{(\Delta,z)}\left(\begin{array}{c}\alpha_n\\ \beta_{n\pm1}\end{array}\right),
\end{equation*}
where $T$ is a time about the same order of magnitude as the signal length, sufficiently long that
the
signal field has effectively decayed to zero everywhere in the medium.
$\hat{U}{(\Delta,z)}$ can be constructed from the operators for free evolution between the pulses
$\hat{U}^{F1}$, $\hat{U}^{F2}$, $\hat{U}^{F3}$ and those for the control pulses 
$\hat{U}^{C1}{(\Delta,z)}$, $\hat{U}^{C2}{(\Delta,z)}$ as
\begin{equation}
 \hat{U}{(\Delta,z)}=\hat{U}^{F3}\cdot\hat{U}^{C2}{(\Delta,z)}\cdot\hat{U}^{F2}
\cdot\hat{U}^{C1}{(\Delta,z)} \cdot\hat{ U}^{F1}
\label{timeevolop}
\end{equation}
Here $\hat{U}^{F1}$, $\hat{U}^{F2}$ and
$\hat{U}^{F3}$, correspond to free evolution during the time intervals $[t_0+T,t_1-T']$,
$[t_1+T',t_2-T']$ and $[t_2+T',t_3-T]$ respectively, 
while $\hat{U}^{C1}{(\Delta,z)}$, $\hat{U}^{C2}{(\Delta,z)}$ to evolution during the
intervals $[t_1-T',t_1+T']$, $[t_2-T',t_2+T']$. The time parameter $T'$ plays the same role as $T$
does for the signal pulse - it is chosen such that the control fields are effectively zero outside
the intervals $[t_j-T',t_j+T']$.
For any $[t,t']$, $\hat{U}^{F}$ is given by  
\begin{equation}
 \hat{U}^{F}=\left(\begin{array}{cc} 1 & 0 \\
0 & e^{-i\Delta(t'-t)}\end{array}\right).
\label{Ufree}
\end{equation}

To construct $\hat{U}^{C1}{(\Delta,z)}$ it is convenient to introduce 
the real envelope and phase functions as $\Omega_1(t)=A_1(t)e^{-i\Phi_1(t)}$ and transform to a
reference frame that rotates with the instantaneous frequency of the pulse using 
\begin{equation*}
 \hat{R}^\dagger_1(t)=\left(\begin{array}{cc}
                  1&0\\0&e^{i\Phi_1(t)}
                 \end{array}\right).
\end{equation*}
Then the equations for $\alpha_r(t)=\alpha(t)$ and $\beta_r(t)=\beta(t)e^{i\Phi_1(t)}$ become:
\begin{equation}
 \partial_t\left(\begin{array}{c} \alpha_r\\\beta_r\end{array}\right)=
i\left(\begin{array}{cc} 0&A_1(t)/2\\A_1(t)/2&\delta(t)\end{array}\right)
\left(\begin{array}{c}\alpha_r\\\beta_r\end{array}\right)
\label{bloch2}
\end{equation}
where we have introduced the instantaneous detuning perceived by the atom 
$\delta(t)=\partial_t\Phi_1(t)-\Delta$.
The Hamiltonian matrix in Eq. \ref{bloch2} can be diagonalized by transforming to the
reference frame of the instantaneous eigenvectors in the standard way \cite{Meystrebook}:
\begin{equation*}
\left(\begin{array}{c} q^+(t) \\ q^-(t) \end{array}\right) =
\hat{V}^\dagger
\left(\begin{array}{c} \alpha_r(t) \\ \beta_r(t) \end{array}\right),
\end{equation*}
where $\hat{V}$ is given by:
\begin{align*}
 \hat{V}&=
\left(\begin{array}{cc} \cos\theta & -\sin\theta \\
\sin\theta & \cos\theta \end{array}\right) & \mathrm{with~~} &
\sin\theta=\frac{A_1}{\sqrt{(\mathcal{R}-\delta)^2+A_1^2}},\\
& & & \cos\theta=\frac{\mathcal{R}-\delta}{\sqrt{(\mathcal{R}-\delta)^2+A_1^2}}\\
& & \mathrm{and~~} & \mathcal{R}=\sqrt{A_1^2+\delta^2}
\end{align*}
Then \ref{bloch2} becomes
\begin{equation}
\partial_t\left(\begin{array}{c} q^+ \\ q^- \end{array}\right) =
i\left(\begin{array}{cc} \lambda^+ & 0 \\
0 & \lambda^- \end{array}\right)
\left(\begin{array}{c} q^+ \\ q^- \end{array}\right)
+(\partial_t\hat{V}^\dagger\cdot\hat{V})
\left(\begin{array}{c} q^+ \\ q^- \end{array}\right)
\label{bloch3}
\end{equation}
where the first term on the RHS contains the adiabatic eigenvalues 
$\lambda^\pm=\frac{1}{2}(\delta\pm\mathcal{R})$
and the second term describes nonadiabatic transitions due to the finite rotation speed of
the basis:
\begin{equation*}
\partial_t\hat{V}^\dagger\cdot\hat{V}=
\left(\begin{array}{cc} 0 & \partial_t\theta \\
-\partial_t\theta & 0 \end{array}\right)
\end{equation*}
If we neglect nonadiabatic transitions, 
we can solve \ref{bloch3} to obtain the time evolution operator in this frame as
\begin{equation}
\hat{U}_{AP}^{C1}=
\left(\begin{array}{cc} e^{i\Lambda_1^+} & 0 \\
0 & e^{i\Lambda_1^-} \end{array}\right) 
\mathrm{~where~} \Lambda_1^\pm=\int_{t_1-T'}^{t_1+T'}\lambda^\pm(t')dt'.
\end{equation}
The $\Lambda_1^\pm$  depend on $\Delta$ through $\lambda^\pm$, as
well as the precise time evolution of $A_1(t)$ and $\partial_t\Phi_1$.

We now assume that the frequency modulation is positive, so $\delta(t_1-T')<0$ and
$\delta(t_1+T')>0$. Then $\sin\theta|_{t_1-T'}=0,\cos\theta|_{t_1-T'}=1$ and
$\sin\theta|_{t_1+T'}=1,\cos\theta|_{t_1+T'}=0$, so in
 the original reference frame we obtain:
\begin{multline}
\hat{U}^{C1}_{(\Delta,z)}=\\
\hat{R}_1(t_1+T')\cdot\hat{V}(t_1+T')\cdot
\hat{U}_{AP}^{C1}\cdot\hat{V}^\dagger(t_1-T')\cdot\hat{R}_1^\dagger(t_1-T')\\
=\left(\begin{array}{cc} 0 & e^{i\Lambda_1^-}e^{i\Phi_1(t_1-T')} \\
-e^{i\Lambda_1^+}e^{-i\Phi_1(t_1+T')} & 0\end{array}\right)
\label{UCP1}
\end{multline}

An identical construction for the second control pulse $\hat{U}^{C2}_{(\Delta,z)}$ and 
a substitution of \ref{Ufree} and \ref{UCP1} into \ref{timeevolop} yields
\begin{multline}
\left[\hat{U}_{(\Delta,z)}\right]_{11}=\\
-e^{i[\Lambda_1^++\Lambda_2^-+\Phi_2(t_2-T')-\Phi_1(t_1+T')]-i\Delta(t_2-t_1-2T')}\\
\shoveleft{\left[\hat{U}_{(\Delta,z)}\right]_{22}=}\\
-e^{i[\Lambda_2^++\Lambda_1^-+\Phi_1(t_1-T')-\Phi_2(t_2+T')]-i\Delta(t_1-t_0+t_3-t_2-2T-2T')}\\
\shoveleft{\left[\hat{U}_{(\Delta,z)}\right]_{12}=\left[\hat{U}_{(\Delta,z)}\right]_{21}=0} 
\label{Umatrix}
\end{multline}
for the matrix elements of $\hat{U}{(\Delta,z)}$ when the conditions for adiabatic passage are
fulfilled.

\bibliography{/home/gdemeter/fiz/pulseprop/publ/bibliography}

\bibliographystyle{apsrev4-1}


\end{document}